\documentclass[preprint,showpacs,aps,superscriptaddress,nobibnotes,ctexart]{revtex4} %
\usepackage{graphicx}
\usepackage{CJK}

\usepackage{bm}
\begin{document}

\title{Effects of dimers on cooperation in the spatial prisoner's dilemma game}
\thanks{This work is supported by the projects of National Natural Science
Foundation of China under Grant No. 10775022, Grant No. 90921015 and
the Fundamental Research Funds for the Central Universities.}
\author{Haihong Li(Àºì)}
\email{haihongli@bupt.edu.cn }  \affiliation {School of Science,
Beijing University of Posts and Telecommunications, Beijing, 100876,
People's Republic of China}
\author{Hongyan Cheng(³ÌºéÑÞ)}
\affiliation {School of Science, Beijing University of Posts and
Telecommunications, Beijing, 100876, People's Republic of China}
\affiliation {Department of Physics, Beijing Normal University,
Beijing, 100875, People's Republic of China}
\author{Qionglin Dai(´úÇíÁÕ)}
\affiliation {School of Science, Beijing University of Posts and
Telecommunications, Beijing, 100876, People's Republic of China}
\author{Ping Ju(¾ÏƼ)}
\affiliation {School of Science, Beijing University of Posts and
Telecommunications, Beijing, 100876, People's Republic of China}
\author{Mei Zhang(ÕÂ÷)}
\affiliation {Department of Physics, Beijing Normal University,
Beijing, 100875, People's Republic of China}
\author{Junzhong Yang(Ñî¿¡ÖÒ)}
\affiliation {School of Science, Beijing University of Posts and
Telecommunications, Beijing, 100876, People's Republic of China}

\begin{abstract}
We investigate the evolutionary prisoner's dilemma game in
structured populations by introducing dimers, which are defined as
that two players in each dimer always hold a same strategy. We find
that influences of dimers on cooperation depend on the type of
dimers and the population structure. For those dimers in which
players interact with each other, the cooperation level increases
with the number of dimers though the cooperation improvement level
depends on the type of network structures. On the other hand, the
dimers, in which there are not mutual interactions, will not do any
good to the cooperation level in a single community, but
interestingly, will improve the cooperation level in a population
with two communities. We explore the relationship between dimers and
self-interactions and find that the effects of dimers are similar to
that of self-interactions. Also, we find that the dimers, which are
established over two communities in a multi-community network, act
as one type of interaction through which information between
communities is communicated by the requirement that two players in a
dimer hold a same strategy.

\end{abstract}

\maketitle

\section{Introduction}
The spontaneous emergence of cooperation in groups of selfish
individuals is ubiquitous in human society and biological systems
and the evolutionary game theory has been considered as an important
approach to investigate the cooperative behavior in those systems.
As one of the most intriguing games, the evolutionary prisoner's
dilemma game (PDG) has attracted much attention over the last few
decades \cite{ax1984} for gaining understanding the emergence of
cooperation. In a PDG, each individual chooses cooperation (C) or
defection (D) as her competing strategy. When the population is
well-mixed, a PDG fails to sustain cooperation, which is often at
odds with reality where mutual cooperation may also be the final
outcome of the game \cite{hof1998, wil308, mili}. In Nowak and May's
seminal works \cite{now359, now1993}, the two-dimensional (2D)
square lattice and the interaction between nearest neighbors enable
cooperators to protect themselves against exploitation of defectors
by forming compact clusters on the lattice. From then on, the
evolutionary PDG on a structured population has been a hot spot. The
influences on cooperation by different factors of models have been
studied intensively. For example, a large set of strategies were
used \cite{jtb192, phy75, prsl257}, different evolutionary rules
were introduced \cite{gab63, mma313}, different types of randomness
were considered \cite{cha405, man433, leb44, gsz58, zhangctp}, and
so on. Also, the influences of population structure on cooperation
have attracted much attention, for example, types of network
structure \cite{now359, Santo95, gomez, szabo07, dura, holme, wu05,
Vukov, wang06, Ohtsuki07, Poncela} and the topological properties of
structure such as average degree \cite{Ohtsuki}, degree-mixing
patterns \cite{zhihai07} and clustering coefficient \cite{Assenza,
cpb09}. Generally, the opinion that the diversity in personalities
of individuals or in population structures could enhance cooperation
in an evolutionary PDG \cite{gab63, wu06, kim02, mas03, svan09,
phys08, njp09, pre046} has been widely accepted.

Now, the investigation on an evolutionary PDG has been extended to
structured populations with communities. Lozano \emph{et al.}
studied the PDG in two practical networks in which inter-community
structure and intra-community structure were designed and found that
cooperation depends strongly on both intra-community heterogeneity
and inter-community connectivity \cite{span1, span2}. In
\cite{njp10}, the authors considered the population with two
communities and studied the effects of inter-connection on
cooperation. They found that cooperation may display a
resonance-like behavior with the variation of the number of
inter-connections.

Consider that, in human society, there always exist many small
coalitions such as family members, colleagues, friends,
collaborators, and so on. The members within one coalition always
hold the same belief and behave in a consistent way. It is an
interesting question that how cooperation varies when the small
coalitions are introduced into structured populations. In this work,
we consider the influences of the factors on cooperation by
introducing dimers to an evolutionary PDG in structured populations.
In each dimer, two players always hold a same strategy. The paper is
organized as follows. In section $2$, the model incorporating dimers
and the categories of dimers are introduced. In section $3$, we
firstly discuss the effects of different categories of dimers on
cooperation in square lattices and ER networks \cite{er}. And then,
we compare the effects of dimers and self-interaction. Finally, we
expand the study to populations with two communities \cite{njp10}
and more rich phenomena are found. In the final section, we give
some discussions and conclusions.

\section{Model}
In a standard evolutionary PDG, there are two steps in one
generation. In the first step, each player follows cooperation
$s_{x}=(\begin{array}{c}
  1 \\
  0 \\
\end{array})$ or defection $s_{x}=(\begin{array}{c}
  0 \\
  1 \\
\end{array})$. The payoff of a player $x$ accumulating by playing PDGs with her
neighbors can be expressed as
\begin{equation}
{P_{x}=\sum_{y\in\Omega_{x}}}s_{x}^{+}Qs_{y},
\end{equation}
where $s_{x}^{+}$ denotes the transpose of the state vector $s_{x}$.
$\Omega_{x}$ includes all of the neighbors of the player $x$. For
simplicity, but without loss of generality, we follow the previous
work \cite{cha405} and adopt the re-scaled payoff matrix $Q$
depending on one single parameter $r$ for PDG
\begin{equation}
{Q=\left(%
\begin{array}{cc}
  1 & -r \\
  1+r & 0 \\
\end{array}%
\right)}, 1<1+r<2.
\end{equation}
In this notation, $1<1+r<2$ measures a defector's temptation to
exploit the neighboring cooperators and $-r$ denotes the sucker's
payoff for a cooperator encountering a defector. Here, $r$ denotes
the ratio of the costs of cooperation to the net benefits of
cooperation. In the second step, the player $x$ will adopt the
strategy $s_{y}$ of a randomly chosen neighbor $y$ with a
probability which is determined by the payoff difference between
them \cite{gsz58}:
\begin{equation}
{W(s_{x}\leftarrow
 s_{y})=1/[{1+exp[(P_{y}-P_{x})/K]}}],
\end{equation}
where the parameter $K$, which is analogous to the temperature in
Fermi-Dirac distribution in statistical physics, characterizes the
stochastic uncertainties in making decisions for the player $x$
\cite{leb44,gsz58}. Throughout the work, we set $K=0.1$.

In this work, the players in the population are divided into two
groups: one with ordinary players who follow the standard
evolutionary PDG, and the other with dimers. The players in dimers
behave differently from the ordinary players only in the step of
strategy updating: in each dimer, the one with higher payoff updates
her strategy ordinarily and the other just follows her partner.
Based on whether the two players in a dimer play game with each
other or not, dimers can be classified into two categories:
interacting ones (I-Dimer) and non-interacting ones (N-Dimer). Each
category of dimers can be subdivided into local dimers (L-Dimer) and
distant dimers (D-Dimer) depending on whether the players in a dimer
are neighbors or not on the given network. To be noted, for
ID-Dimer, the given network structure is modified by extra
connections between players in dimers and, for NL-Dimer, the given
network structure is modified by cutting the connections between
players in dimers. In this work, we do not consider NL-Dimer and
just focus on the effects of the other three categories of dimers on
cooperation in different types of population structures such as
square lattices with periodic boundary conditions and degree $z=4$,
Erd\"{o}s-R\'{e}nyi (ER) \cite{er} networks with mean degree $z=8$
and structural populations with two communities. In a
two-community-structure population where one community is a square
lattice with $z=4$ and the other a square lattice with $z=8$, dimers
are established over these two communities, that is, two players in
a dimer locate on different communities.

Throughout this work, we set the number of players in the population
to be $N=10,000$. Initially, in the Monte Carlo simulations, $m$
dimers are randomly assigned through the population and players take
the strategy of C or D with equal probability. To measure the
cooperation level, the cooperator frequency $\rho_{c}$ will be
monitored when the evolution of strategy pattern reaches its steady
state. All the following data are obtained with synchronous strategy
updating and each point is gained by averaging $1000$ generations
after a transient time of $6000$ generations and by averaging over
$100$ independent realizations.

\section{Simulation results and analysis}

\begin{figure*}
\includegraphics[width=5in]{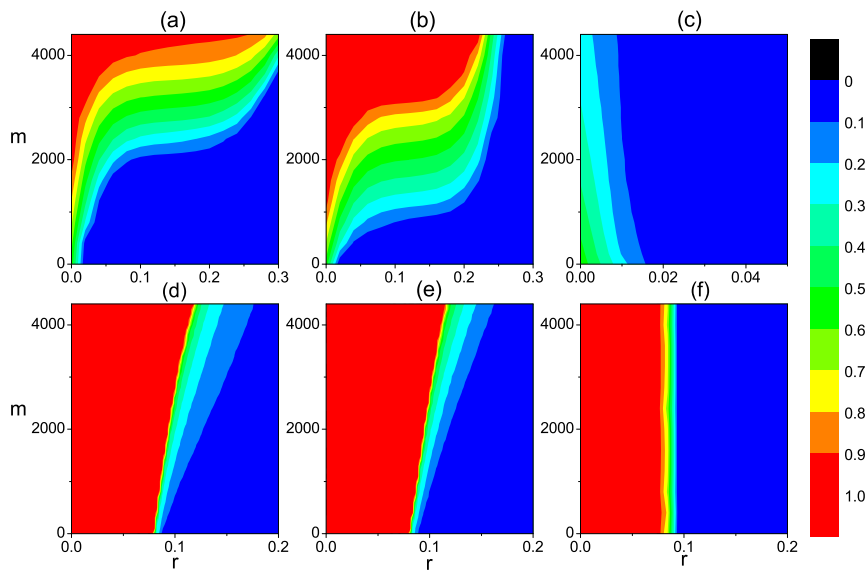}
\caption{\label{fig1} (color online) The contour graphs of the
cooperation level $\rho_{c}$ for evolutionary PDGs, as functions of
$r$ and the number of dimers $m$ on square lattices with $z=4$ (top
panel) and ER networks with mean degree $z=8$ (bottom panel). $m$
increases from $0$ to $4400$ in all cases. (a, d) For IL-Dimer. (b,
e) For ID-Dimer. (c, f) For ND-Dimer.}
\end{figure*}

Generally, cooperation is maintained in an evolutionary PDG on
networks by forming cooperator clusters (C-clusters). The
interactions between cooperators at the boundaries of C-clusters and
those inside C-clusters enable cooperators at the boundaries to have
high payoffs to compete with surrounding defectors. Intuitively, the
same strategy held by the players in a dimer make it possible that
cooperator dimers (C-dimer) serve as seeds for C-clusters and tend
to accelerate the expansion of C-clusters. Therefore, it seems that
the presence of dimers in an evolutionary PDG would improve
cooperation. However, as we show below, the effects of dimers on
cooperation are not self-evident and whether cooperation is improved
or not depends on the type of dimers and the structure of the
underlying networks.

Firstly, we investigate the effects of dimers on cooperation in
populations without multi-community.  The contour graphs of
cooperator frequency $\rho_{c}$ as functions of $r$ and the number
of dimers $m$ are presented in figure~1. The top panel shows the
results for IL-Dimer, ID-Dimer and ND-Dimer on square lattices,
respectively. For the range of $r$ that is not covered here, almost
all of $\rho_{c}$ reaches $0$ or will descend to $0$ and no more
raise of $\rho_{c}$ appears. Interestingly, the presence of
interaction between players in a dimer plays a decisive role on
cooperation. Dimers enhance cooperation strongly for both IL-Dimer
and ID-Dimer. Especially, in these two situations, the state that
all players become cooperators (All-C state) may be reached for a
large range of $r$ where cooperators die off in the absence of
dimers. Furthermore, ID-Dimer shows a faster growth of $\rho_{c}$
with the number of dimers than IL-Dimer in the range of $r<0.2$,
which indicates a stronger cooperation enhancement for ID-Dimer than
IL-Dimer. The stronger cooperation enhancement for ID-Dimer results
from the presence of shortcuts established by the two players in
each dimer. Together with a same strategy held by players in a
dimer, these shortcuts shorten the average distance between any two
players, which speeds up the expansion of C-clusters. On the other
hand, the presence of ND-Dimer deteriorates cooperation, i.e.,
$\rho_{c}$ decreases with the number of dimers. The deterioration of
cooperation for ND-Dimer could be explained by the behavior of
C-dimer. Since there are no interactions between players in these
C-dimers, the advantage in payoff for cooperators at the boundaries
of C-clusters over defectors may be weakened. Additionally, due to
the absence of interaction between the players in these C-dimers,
the speeding up of the expansion of C-clusters is lost and the
defectors will benefit from these C-dimers provided that they are
the neighbors of these C-dimers. Both of these suppress cooperation
in ND-Dimer and the suppression of cooperation increases with the
number of dimers.

The bottom panel in figure~1 shows the results for IL-Dimer,
ID-Dimer and ND-Dimer on ER networks, respectively. In comparison
with the top panel in figure~1, the improvement of cooperation by
IL-Dimer and ID-Dimer is observed, though the improvement is not
prominent. However, the influence of ND-Dimer on cooperation is
quite different from that on square lattices: $\rho_{c}$ is
insensitive to the presence of dimers in this situation. Consider
that, in ER networks where the mean distance between any two players
is short, the players in C-clusters are always exposed to defectors
and defectors always benefit from cooperators in C-clusters in an
ordinary PDG. When ND-Dimer is introduced, defectors cannot get more
benefits from C-dimer than those in the absence of ND-Dimer.
Therefore, the deterioration of cooperation by ND-Dimer on square
lattices is missed on ER networks and cooperation is independent of
the number of dimers.

\begin{figure*}
\includegraphics[width=4.5in]{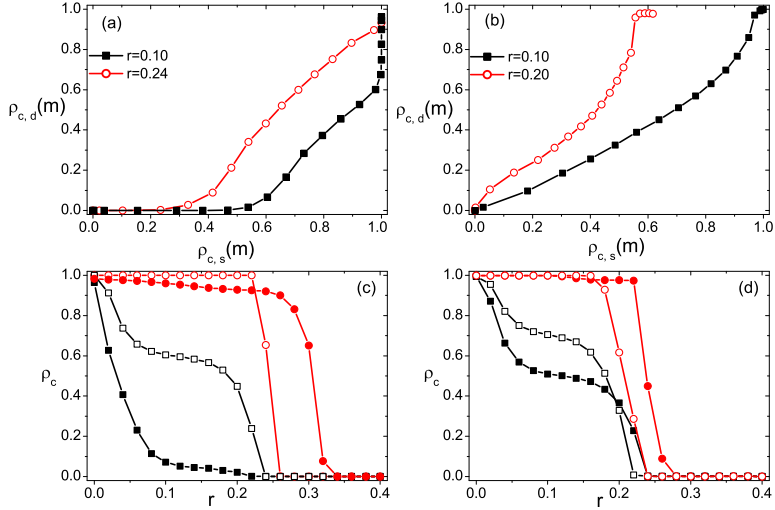}
\caption{\label{fig2} (color online) The relationships between the
cooperator frequencies $\rho_{c, s}$ for square lattices with
self-interaction and $\rho_{c, d}$ for those with dimers. (a)
$\rho_{c, d}$ is for IL-Dimer. Closed squares are for $r=0.10$ and
open circles for $r=0.24$. (b) $\rho_{c, d}$ is for ID-Dimer. Closed
squares are for $r=0.10$ and open circles for $r=0.20$. (c) The
closed symbols are for IL-Dimer with $m=2000$ (squares) and $m=4400$
(circles). The open symbols are for the case without dimers and in
which there are $4000$ (squares) or $8800$ (circles) players with
self-interaction. (d) The closed symbols are for ID-Dimer with
$m=2000$ (squares) and $m=4400$ (circles). The open symbols are for
the case without dimers and in which there are $4000$ (squares) or
$8800$ (circles) players with self-interaction. The structure of
networks for the case with self-interaction is same as that in
ID-Dimer, which is modified by the shortcuts between players in
dimers. }
\end{figure*}

It is well known that the inclusion of self-interaction in an
evolutionary PDG on networks \cite{gsz58} favors cooperation.
However, how self-interaction relates to reality is unknown. As far
as it is concerned that a same strategy is held by two players in a
dimer, we find that the interaction between players in a dimer for
IL-Dimer or ID-Dimer acts as self-interaction and either IL-Dimer or
ID-Dimer provides a way to realize self-interaction for players to
some extent. To make it clear, we compare the effects of IL-Dimer
(or ID-Dimer) and self-interaction on cooperation. We take square
lattices with $z=4$ as examples. It has to be mentioned that the
square lattices with the presence of ID-Dimer are modified by the
shortcuts between players in dimers. Therefore, the networks with
self-interaction to be studied are 'square lattices' with the same
shortcuts as those established by dimers. To be noted, not all
players on networks have self-interaction and the number of players
with self-interaction is twice as many as the number of dimers. For
a given number of dimers $m$, we monitor the cooperator frequencies
$\rho_{c,s}(m)$ for networks with self-interaction and
$\rho_{c,d}(m)$ for networks with dimers. As shown in the top panel
in figure~2 where the relationships between $\rho_{c,s}(m)$ and
$\rho_{c,d}(m)$ are presented, $\rho_{c,s}(m)$ is positively
correlated with $\rho_{c,d}(m)$ and the slope of $\rho_{c,d}$ over
$\rho_{c,s}$ is roughly around $1$, which means that IL-Dimer or
ID-Dimer does provide a way to realize self-interaction. However,
some differences exist between the systems with self-interaction and
those with IL-Dimer or ID-Dimer. For example, as shown in the bottom
panel in figure~2 where $\rho_{c,s}$ and $\rho_{c,d}$ against $r$
for different $m$ are presented, $\rho_{c,s}$ is always higher than
$\rho_{c,d}$ for small $r$ whereas $\rho_{c,s}$ becomes lower than
$\rho_{c,d}$ for large $r$.

Additionally, the above model is investigated in square lattices and
ER networks with some other mean degrees, such as $z=8$ for square
lattices and $z=4$ for ER networks. The analogous phenomena could be
obtained. Since that there is an explicit dependence between the
value of the $K$ and the outcome of the prisoner's dilemma following
the Fermi update rule, we test the work with different $K$ and find
the robustness of our results against the variation of $K$.

\begin{figure*}
\includegraphics[width=5in]{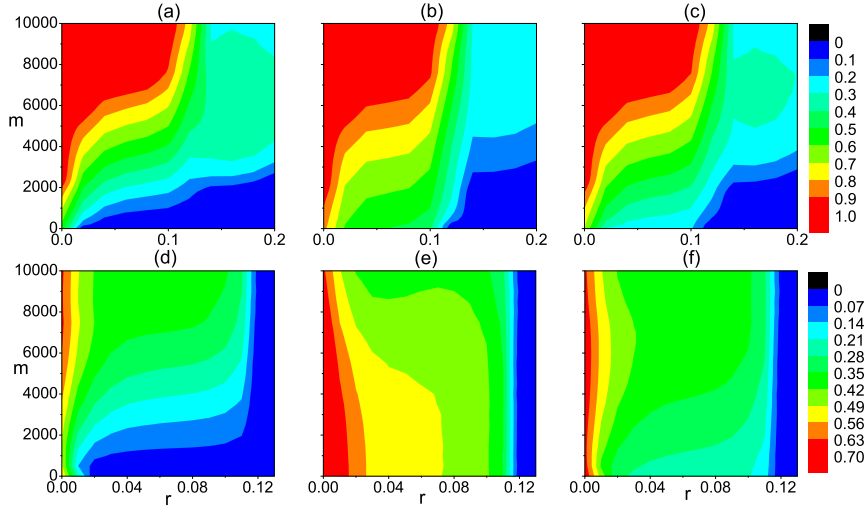}
\caption{\label{fig3} (color online) The contour graphs of the
cooperation level $\rho_{c}$, as a function of $r$ and the number of
dimers $m$ in two communities with random dimer. The top panel is
for ID-dimer and the bottom one is for ND-dimer. (a, d) $z=4$. (b,
e) $z=8$. (c, f) For the averaged $\rho_{c}$ over two communities. }
\end{figure*}

Secondly, we consider the effects of dimers on a population with two
communities: one is a square lattice with $z=4$ and the other with
$z=8$. As stated in the model section, two players in each dimer
belong to different communities and are randomly selected. Clearly,
dimers on this population structure fall into the category of either
ID-Dimer or ND-Dimer. The effects of dimers on cooperation for
ID-Dimer and ND-Dimer are illustrated in the top panel and the
bottom panel in figure~3, respectively. The left and middle columns
show the contour graphs of $\rho_{c}$ on the $r$-$m$ space for the
two communities respectively and the right column is for the whole
population. Similar to the populations without community structure,
cooperation is improved for ID-Dimer and a high level of cooperation
may be reached. A resonance-like behavior with the variation of the
number of dimers could be observed when $r>0.1$, which is similar to
that in \cite{njp10} which was discussed particularly. In contrast
with the populations without community structure, ND-Dimer indeed
enhances cooperation. For ND-Dimer, though cooperation is a little
downgraded in the community with $z=8$ in which $\rho_{c}$ is
higher, cooperation, both in the whole population and in the
community with $z=4$ in which $\rho_{c}$ is lower, is enhanced in
comparison with that in the absence of dimers. As mentioned above,
the deterioration of cooperation in a square lattice with ND-Dimer
originates from two factors involving C-dimers: players in C-dimers
locating inside C-clusters can not support their partners at the
boundaries of C-clusters, yet support the defectors neighboring
their partners indirectly. However, for ND-Dimer defined over two
communities, the deterioration of cooperation by the first factor is
lost and, consequently, the downgrade of cooperation by ND-Dimer in
the community $z=8$ with a high level of cooperation becomes weak.
On the other hand, the requirement that the players in dimers hold a
same strategy may enhance cooperation if one of players is in a
C-cluster. Combining these together, we observe a weak deterioration
of cooperation in one community whereas the enhancement of
cooperation in the other one and in the whole population.

An interesting comparison can be made between the results in this
work and those in \cite{njp10}. Both works consider the evolution of
cooperation in a population with two interacting communities, but
the ways of interaction are different. Either interaction through
ID-Dimer or interaction through ND-Dimer may lead to two different
evolutions of cooperation in comparison with those in \cite{njp10}.
For interaction with type of ID-Dimer, interaction between
communities always improves cooperation, which is independent of the
cooperation levels in isolated communities. Actually, even when
cooperation is extinct in both isolated communities, cooperation may
still reach a high level. For interaction with type of ND-Dimer, the
cooperation levels in two communities may decrease simultaneously
for weak interaction, and a resonance-like behavior of cooperation
against interaction strength may appear for that the cooperation
levels in two communities are not in the intermediate range.

\section{Conclusions}
In conclusion, we introduce dimers in which two players hold the
same strategy to an evolutionary PDG on structured populations. The
effects of dimers on cooperation are investigated. We find that
influences of dimers on cooperation depend on the type of dimers and
the population structure. For example, ID-Dimer always enhances
cooperation and a high level of cooperation may be reached with a
large number of dimers. However, the influences of ND-Dimer on
cooperation depend on the population structure strongly:
deterioration of cooperation on square lattices, little variation of
cooperation on ER networks, and enhancement of cooperation on the
population with two communities. Some interesting discussions
between the results in this work and previous studies are made.
Firstly, we discuss the relationship between dimers and
self-interactions and find that the effect of dimers is similar to
that of self-interaction. Secondly, we compare the interaction
through interconnections between communities, where interaction is
realized by playing games between players in different communities,
with that through dimers, where interaction is realized through
holding a same strategy by two players in a dimer. For the
interaction through dimers, rich phenomena are observed.

\section*{References}


\end{document}